\documentclass[apj,iop]{emulateapj}
\usepackage{amsmath}
\usepackage{graphicx}
\usepackage{color}

\shorttitle{Transverse oscillations of loops with coronal rain observed by \textit{Hinode}/SOT}
\shortauthors{P. Antolin \& E. Verwichte}

\begin{document}

\title{Transverse oscillations of loops with coronal rain observed by \textit{Hinode}/SOT}

\author{P. Antolin\altaffilmark{1,3} and E. Verwichte\altaffilmark{2}}
\affil{\altaffilmark{1}Institute of Theoretical Astrophysics, University of Oslo, P.O. Box 1029, Blindern, NO-0315 Oslo, Norway}
\affil{\altaffilmark{2}Department of Physics, University of Warwick, Coventry CV4 7AL, UK}
\altaffiltext{3}{Also at: Center of Mathematics for Applications, University of Oslo, P.O. Box 1053, Blindern, NO-0316, Oslo, Norway}
\email{patrick.antolin@astro.uio.no, erwin.verwichte@warwick.ac.uk}

\begin{abstract}

The condensations composing coronal rain, falling down along loop-like structures observed in cool chromospheric lines such as H$\alpha$ and \ion{Ca}{2} H, have long been a spectacular phenomenon of the solar corona. However, considered a peculiar sporadic phenomenon, it has not received much attention. This picture is rapidly changing due to recent high resolution observations with instruments such as \textit{Hinode}/SOT, CRISP of \textit{SST} and \textit{SDO}. Furthermore, numerical simulations have shown that coronal rain is a loss of thermal equilibrium of loops linked to footpoint heating. This result has highlighted the importance that coronal rain can play in the field of coronal heating. In this work, we further stress the importance of coronal rain by showing the role it can play in the understanding of the coronal magnetic field topology. We analyze \textit{Hinode}/SOT observations in the \ion{Ca}{2} H line of a loop in which coronal rain puts in evidence in-phase transverse oscillations of multiple strand-like structures. The periods, amplitudes, transverse velocities and phase velocities are calculated, allowing an estimation of the energy flux of the wave and the coronal magnetic field inside the loop through means of coronal seismology. We discuss the possible interpretations of the wave as either standing or propagating torsional Alfv\'en or fast kink waves. An estimate of the plasma beta parameter of the condensations indicates a condition that may allow the often observed separation and elongation processes of the condensations. We also show that the wave pressure from the transverse wave can be responsible for the observed low downward acceleration of coronal rain.

\end{abstract}


\section{Introduction}

There is now increasing evidence that active regions in the Sun have their heating concentrated mostly at lower atmospheric regions, from the lower chromosphere to the lower corona. The excess densities found in most observed coronal structures such as coronal loops put them out of hydrostatic equilibrium, a state that can be explained by footpoint heating \citep{Aschwanden_2001ApJ...560.1035A}. \citet{Hara_2008ApJ...678L..67H} using the \textit{Hinode}/EIS instrument, have shown that active region loops exhibit upflow motions and enhanced nonthermal velocities at their footpoints. The characteristic over-density has also been deduced seismologically by \citet{VanDoorsselaere_etal_2007AA...473..959V} by studying the fundamental to the second harmonic period ratio $P_{1}/P_{2}$ of standing transverse oscillations in loops. Recently, \citet{DePontieu_etal_2011Sci...331...55D} have highlighted the importance of the link between the photosphere and the corona by showing that a considerable part of the hot coronal plasma could be heated at low spicular heights, thus explaining the fading character of the ubiquitous 'type II spicules'. Further evidence of footpoint heating is put forward by the presence of cool structures in the active region coronae, such as filaments/prominences or coronal rain, two phenomena that may share the same formation mechanism but which seem to differ on the structure of their underlying magnetic field topology, leading to different observational aspects such as dynamics, shapes and lifetimes. 

Both, prominences (or filaments if observed on disk rather than at the limb) and coronal rain correspond to cool and dense plasma observed at coronal heights in chromospheric lines such as H$\alpha$ and \ion{Ca}{2} H and K. But while the plasma in prominences is suspended in the corona against gravity making the structures long-lived (days to weeks), coronal rain is observed falling down in timescales of minutes \citep{Schrijver_2001SoPh..198..325S, DeGroof_2004AA...415.1141D} along curved loop-like trajectories. The mechanical stability and thermodynamic properties of prominences are linked with the underlying magnetic field topology, and thus the main difference between coronal rain and prominences could be a difference in coronal magnetic field configuration. This question awaits further proper investigation.

While prominences have been studied extensively in solar physics, few observational studies exist of coronal rain since their discovery in the early 1970s \citep{Kawaguchi_1970PASJ...22..405K, Leroy_1972SoPh...25..413L}. This lead to the belief that coronal rain is a rather uncommon phenomenon in active region coronae. Furthermore, coronal rain is often erroneously attributed to prominence material falling back from coronal heights following a prominence eruption. However, recent high resolution observations with instruments such as CRISP of \textit{SST}, SOT of \textit{Hinode} or \textit{SDO} reveal coronal rain to be dynamic, short-lived (1-10 minutes), small sized (200 km or less) strand-like structures that are ubiquitously present over active regions \citep{Antolin_etal_11}. The detection of coronal rain therefore requires high spatial and temporal resolution observations that have only recently become available. The same type of observations have shown that prominences are composed of a myriad of fine threads, outlining a fine-scale structure of the magnetic field, and the presence of flows along the threads \citep{Heinzel_Anzer_2006ApJ...643L..65H, Lin_2005SoPh..226..239L, Lin_2008ASPC..383..235L, Lin_2010SSRv..tmp..112L,Martin_2008SoPh..250...31M}. The frequency of coronal rain has profound implications for coronal heating \citep{Antolin_2010ApJ...716..154A}. In order to quantify the true occurrence frequency and importance of coronal rain in active region loops we will require further effort gathering a statistically significant number of high resolution observations in different wavelengths. 

Numerical simulations have shown that coronal rain and prominences are most likely the result of a phenomenon of thermal instability, also known as 'catastrophic cooling' \citep{Hildner_1974SoPh...35..123H, Antiochos_1999ApJ...512..985A, Muller_2003AA...411..605M, Muller_2004AA...424..289M, Mendozabriceno_2005ApJ...624.1080M, Antolin_2010ApJ...716..154A}. Loops with footpoint heating present high coronal densities and thermal conduction turns out to be insufficient to maintain a steady heating per unit mass, leading to a gradual decrease of the coronal temperature. Eventually recombination of atoms takes place and temperature decreases to chromospheric values abruptly in a timescale of minutes locally in the corona. This is accompanied by local pressure losses leading to the formation of condensations which become bright or dark if observed towards the limb or on disk, respectively.

\citet{Antolin_2010ApJ...716..154A} showed that Alfv\'en wave heating, a strong coronal heating candidate, is not a predominant heating mechanism in loops with coronal rain. When propagating from the photosphere into the corona, Alfv\'en waves can nonlinearly convert to longitudinal modes through mode conversion due to density fluctuations, wave-to-wave interaction, and deformation of the wave shape during propagation. These modes subsequently steep into shocks and heat the plasma uniformly along the loop \citep{Moriyasu_2004ApJ...601L.107M, Antolin_2010ApJ...712..494A, Vasheghani_2011AA...526A..80V}, thus avoiding the loss of thermal equilibrium in the corona. Coronal rain is often observed falling down at speeds much lower than free fall speeds resulting from the effective gravity along loops \citep{Schrijver_2001SoPh..198..325S, DeGroof_2004AA...415.1141D, DeGroof05, Muller_2005AA...436.1067M, Antolin_2010ApJ...716..154A}. Simulations have shown that the effects of gas and magnetic pressure may explain the observed dynamics \citep{Mackay_2001SoPh..198..289M, Muller_2003AA...411..605M, Antolin_2010ApJ...716..154A}. Here we show that the observed wave pressure from a transverse wave may also account for decreased accelerations.

Magnetohydrodynamic (MHD) waves are frequently observed in prominences \citep{Ramsey_1966AJ.....71..197R, Oliver_Ballester_2002SoPh..206...45O, Foullon_2004AA...427L...5F, Lin_etal_2007SoPh..246...65L, Lin_2009ApJ...704..870L, Okamoto_2007Sci...318.1577O, Terradas_etal_2008ApJ...678L.153T} and coronal loop structures \citep{Aschwanden_1999ApJ...520..880A, DeMoortel_2000AA...355L..23D, VanDoorsselaere_etal_2008AA...487L..17V, Erdelyi_2008AA...489L..49E, Verwichte_2010ApJ...717..458V}, leading to the determination of the internal physical conditions through the development of analytical theory and numerical modeling \citep{Roberts_1984ApJ...279..857R, Nakariakov_2005LRSP....2....3N, Ballester_2006RSPTA.364..405B, Andries_2009SSRv..149....3A, Taroyan_2009SSRv..tmp...24T, Arregui_Ballester_2010arXiv1011.5175A}, a technique dubbed coronal seismology. The determination of the physical properties of the corona through which MHD waves travel depends on the correct interpretation of the observed signatures, correct identification of the wave mode and an MHD wave model that provides robust seismological measurements. For waves in transients objects such as spicules and filament fibres, the role of a wave guide has been debated. 

\citet{Okamoto_2007Sci...318.1577O} analyzed transverse oscillations running through prominence threads observed by \textit{Hinode}/SOT in the \ion{Ca}{2} H-line at the limb of the Sun. 
The reported mean periods for the waves are between 130 s and 240 s, (horizontal) oscillation amplitudes between 400 km and 1770 km, transverse (vertical) velocities between 5 km s$^{-1}$ and 15 km s$^{-1}$ and an estimated wave speed larger than 1050~km~s$^{-1}$ leading to a magnetic field of $50$ G in the prominence. Minimum Alfv\'en speeds in the prominence were estimated by \citet{Terradas_etal_2008ApJ...678L.153T} to be between 120~km~s$^{-1}$ and 350~km~s$^{-1}$, depending on the local magnetic field, the total lengths of the magnetic field lines in the prominence, and the ratio between the local and external (coronal) density. \citet{Okamoto_2007Sci...318.1577O} first interpreted the oscillations as Alfv\'en waves running through the prominence. However, \citet{Terradas_etal_2008ApJ...678L.153T} and \citet{VanDoorsselaere_2008ApJ...676L..73V} have argued that the only solution among fast waves that gives rise to a displacement of a magnetic flux tube axis is the kink mode. Furthermore, \citet{Terradas_etal_2008ApJ...678L.153T} have shown that the periods of the kink mode are rather insensitive to the presence of steady flows along the threads.

\citet{Ofman_Wang_2008AA...482L...9O} have analyzed an event with \textit{Hinode}/SOT in the \ion{Ca}{2} H line, similar to the one in the present work. Transverse oscillations are observed in a loop with flows. In this case, the cool material is ejected at very high speeds ($74-123$~km~s$^{-1}$) from one footpoint to the other, and is related to a flare happening close-by, which may also be the cause for the oscillations. The waves are interpreted mostly as fundamental modes of standing kink oscillations, although some of the observed threads display dynamics more consistent with propagating fast magnetoacoustic waves. Coronal seismology is performed assuming a density in the range $(1-5)\times10^{9}$~cm$^{-3}$, leading to coronal magnetic fields of $20\pm7$~G. The analyzed loop does not seem to be subject to catastrophic cooling during the observed time, and thus the flow is of a different nature than that of the present work. The cause for the observed oscillations in our case seems to be different as well, since no energetic phenomenon is observed.

In this work we analyze the same high resolution observations of \citet{Okamoto_2007Sci...318.1577O}, but concentrate on active region loops in the foreground, unconnected to the prominence, that exhibit coronal rain. In \cite{Antolin_2010ApJ...716..154A} the observational analysis concentrated on loops to the north of the visible sunspot. Here, we will focus on one loop on the south side of this sunspot and which exhibits a peculiar phenomenon. We present the first observational analysis of transverse oscillations of threads in a loop subject to coronal rain. The paper is organized as follows. In Section \ref{observations} we describe the data set of \textit{Hinode}/SOT, present statistics of velocities and accelerations for the falling coronal rain and analyze the observed oscillations in the loop. In Section \ref{discussion} we proceed to discuss the observational results, giving interpretations for the wave properties and their nature, and finalize in Section \ref{conclusions} with the conclusions of the work.

\begin{figure*}
\epsscale{1.}
\plotone{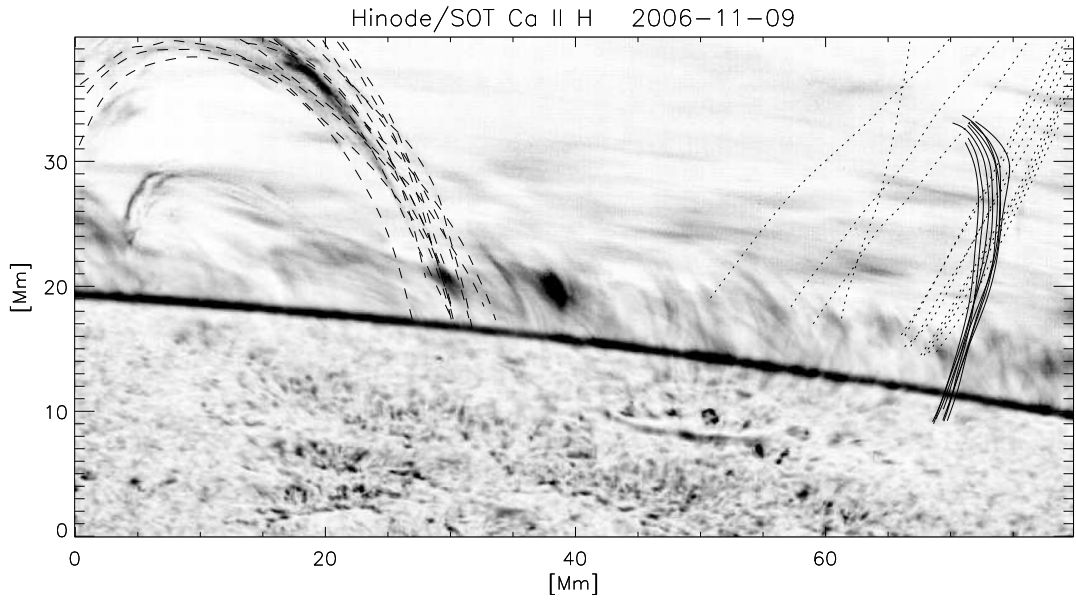}
\caption{Active Region NOAA 10921 on the west limb observed by \textit{Hinode}/SOT in the \ion{Ca}{2}~H band on the 9 November 2006 between 19:33 and 20:44 UT. The curves denote some of the paths traced by coronal rain. The solid curves conform the loop studied in the present work, while the loops outlined by the dashed curves were studied in \citet{Antolin_2010ApJ...716..154A}. The dotted curves mark the presence of other loops which may be interacting with the loop studied here.
\label{fig0}}
\end{figure*}

\begin{figure*}
\epsscale{1.}
\hspace*{-0.4in}\plotone{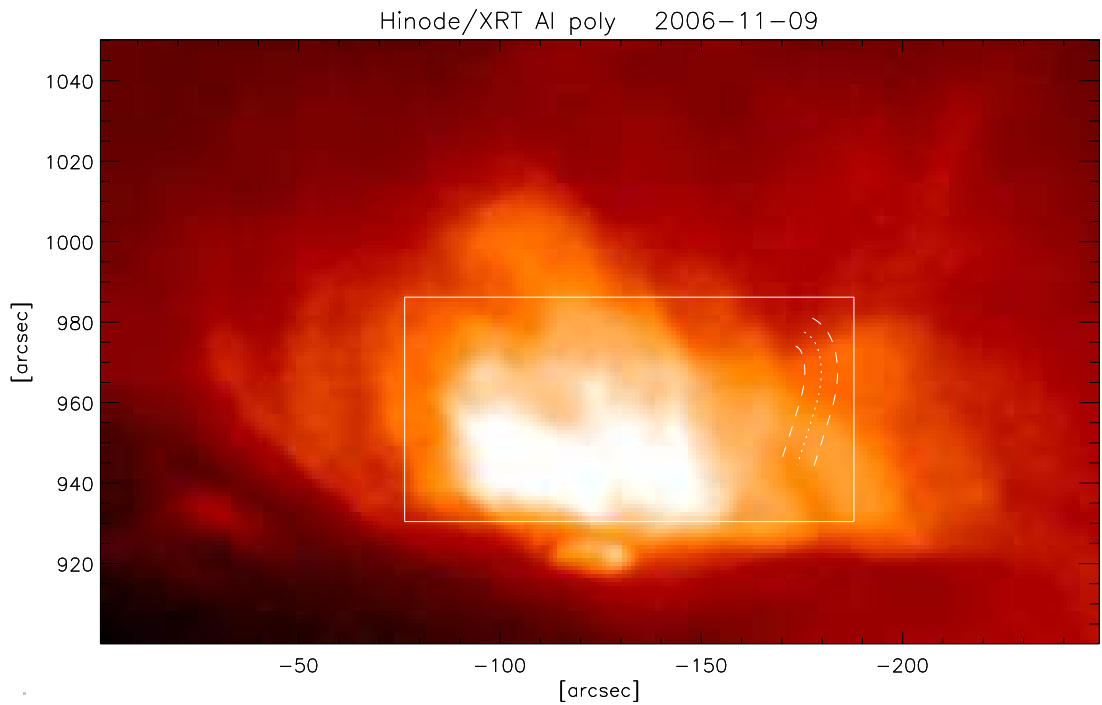}
\caption{Active Region NOAA 10921 observed by \textit{Hinode}/XRT with the Al Poly filter on the 9 November 2006 at 19:59 UT, roughly half an hour before the observed coronal rain in the studied coronal loop. The solid and dashed lines mark the position of the loop, as in Figure \ref{fig2}.
\label{fig1}}
\end{figure*}

\begin{figure}
\hspace*{-0.2in}\epsscale{1.3}
\plotone{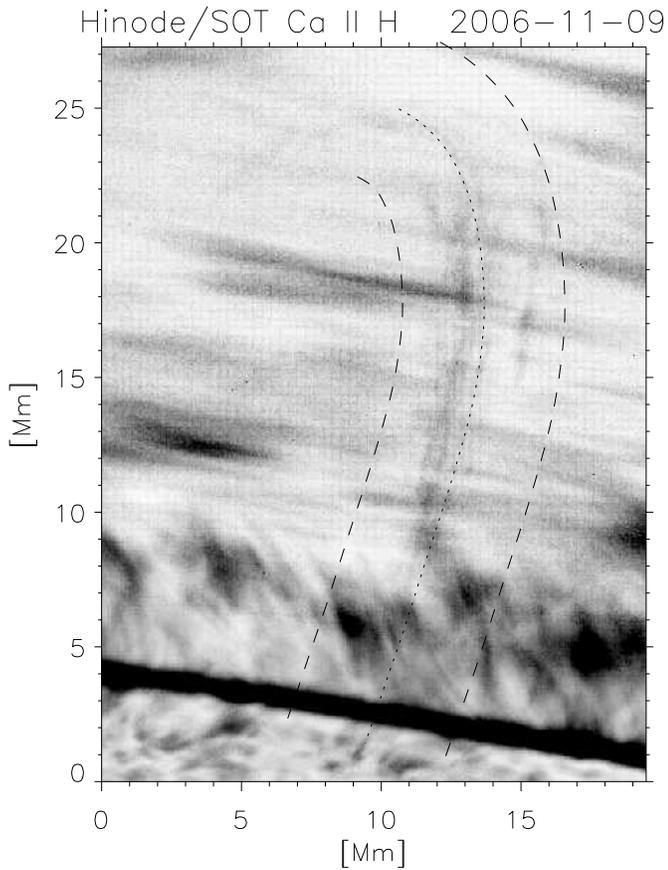}
\caption{Loop with coronal rain observed by \textit{Hinode}/SOT on the 9 November 2006 above AR 10921.
\label{fig2}}
\end{figure}

\section{Observations of coronal rain with \textit{Hinode}/SOT}\label{observations}

\subsection{Velocities and accelerations}

The observations with the Solar Optical Telescope (SOT) of \textit{Hinode} \citep{Tsuneta_2008SoPh..249..167T} are in the \ion{Ca}{2} H band, on 9 November 2006 from 19:33 to 20:44 UT with a cadence of 15 s and a spatial resolution of $1.22\lambda/D\simeq0.\!\!^{\prime\prime}$2, and focused on NOAA AR 10921 on the west limb. A variance of the images over a part of the time interval is shown in Figure~\ref{fig0}. This data set has become famous among solar limb observations. The set shows the presence of an active region prominence which exhibits interesting oscillatory behaviour \citep{Okamoto_2007Sci...318.1577O}. Figure \ref{fig1} shows a \textit{Hinode}/XRT observation of the same region at 19:59 UT with the Al Poly filter. The square in the Figure corresponds to the SOT field of view.

Additionally, on the foreground of the prominence various loops exhibiting coronal rain have been observed. A statistical study of the loops outlined in dashed curves in Figure~\ref{fig0} can be found in \citet{Antolin_2010ApJ...716..154A}. The observed loops are located north of the sunspot, have lengths between 60 Mm and 100 Mm, and exhibit coronal rain continuously. The condensations composing coronal rain display a broad distribution of velocities (between $20$ km s$^{-1}$ and $120$ km s$^{-1}$) that put in evidence both acceleration and deceleration processes in the loops. The average accelerations were found to be lower than that produced by gravity, indicating the presence of other forces, possibly of magnetic origin.

The focus of our paper is set on a coronal loop located south of the sunspot, outlined in solid curves in Figure~\ref{fig0}, which can be observed in the \ion{Ca}{2} H band of \textit{Hinode}/SOT thanks to the coronal rain occurring in the loop. Figure \ref{fig2} shows the subset of the entire field of view corresponding to this loop. The loop is visible for about half an hour towards the end of the observation set. We have plotted the variance of the image over the period of time it becomes visible. Assuming that the geometry of the loop is close to that of a semi-torus, we see from Figure \ref{fig2} that the plane of the loop makes a significant angle with the plane of the sky, being roughly directed along the line of sight and that it is slightly inclined with respect to the vertical. The coronal rain can be observed basically from the apex of the loop, located $25\pm5$ Mm above the surface, leading to a loop length of $80\pm15$ Mm assuming a circular axis for the loop. In the \textit{Hinode}/XRT image of Figure \ref{fig1} we have outlined the position of this loop.

We have tracked down the condensations along the loop with the help of CRISPEX (CRisp SPectral EXplorer) and Timeslice ANAlysis Tool (TANAT) \footnote{The actual code and further information can be found at \\ http://www.astro.uio.no/~gregal/crispex/index.html}, two widget based tools programmed in the Interactive Data Language (IDL), which enable the easy browsing of the image (and if present, also spectral) data, the determination of loop paths, extraction and further analysis of length-time diagrams. 

\begin{figure}[top]
\begin{center}
$\begin{array}{c}
\includegraphics[scale=0.45]{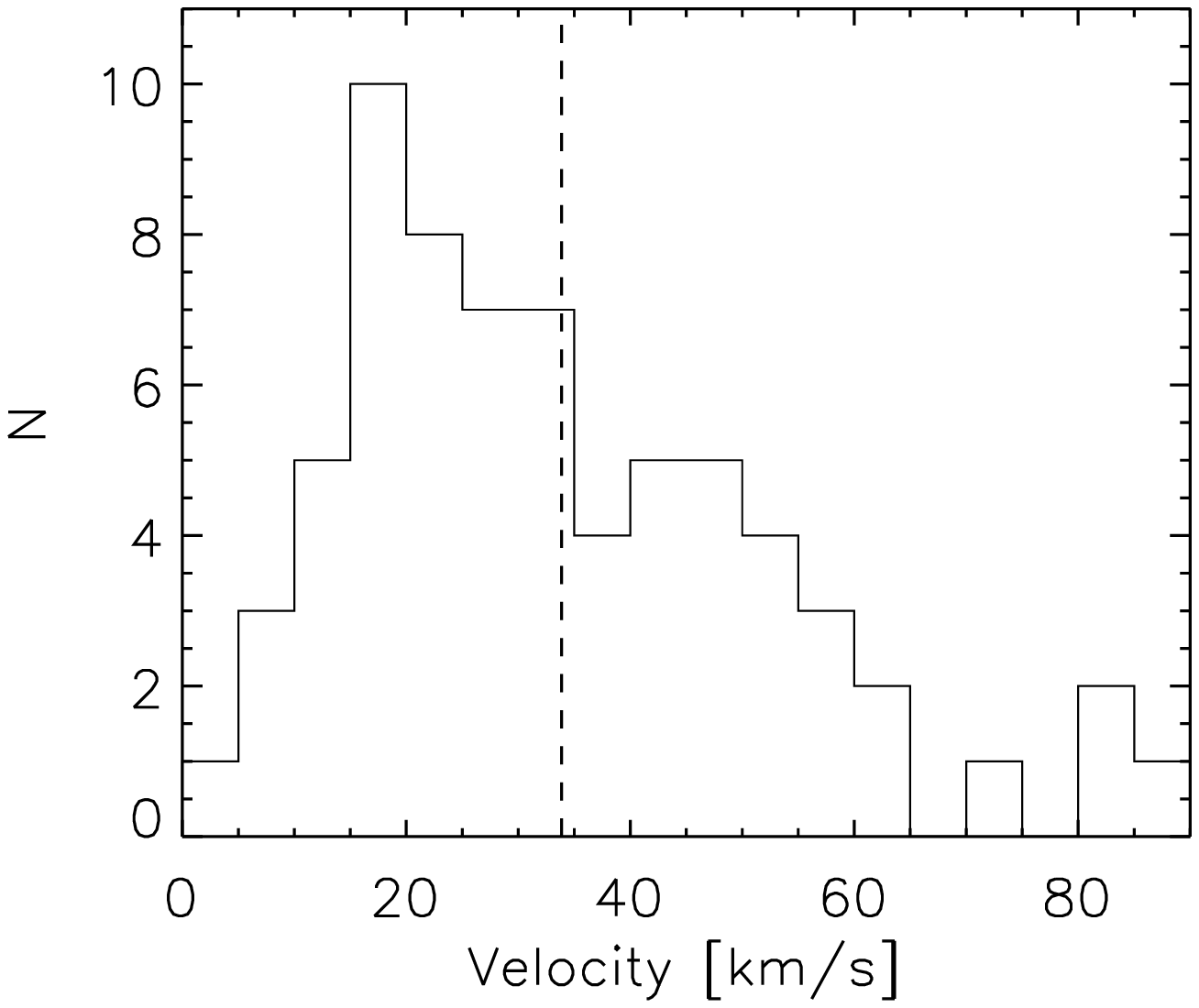}  \\
\includegraphics[scale=0.45]{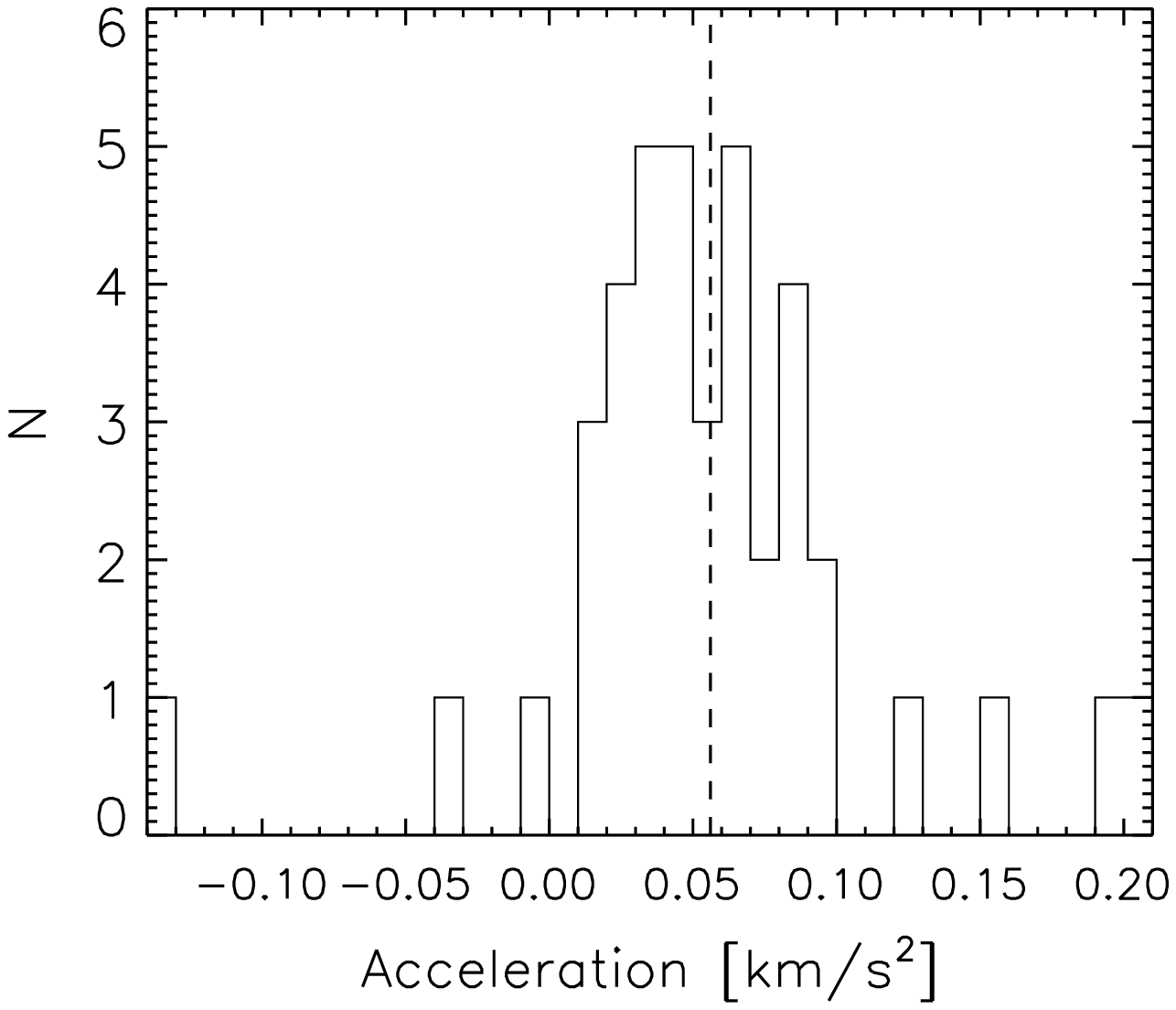} \\
\includegraphics[scale=0.45]{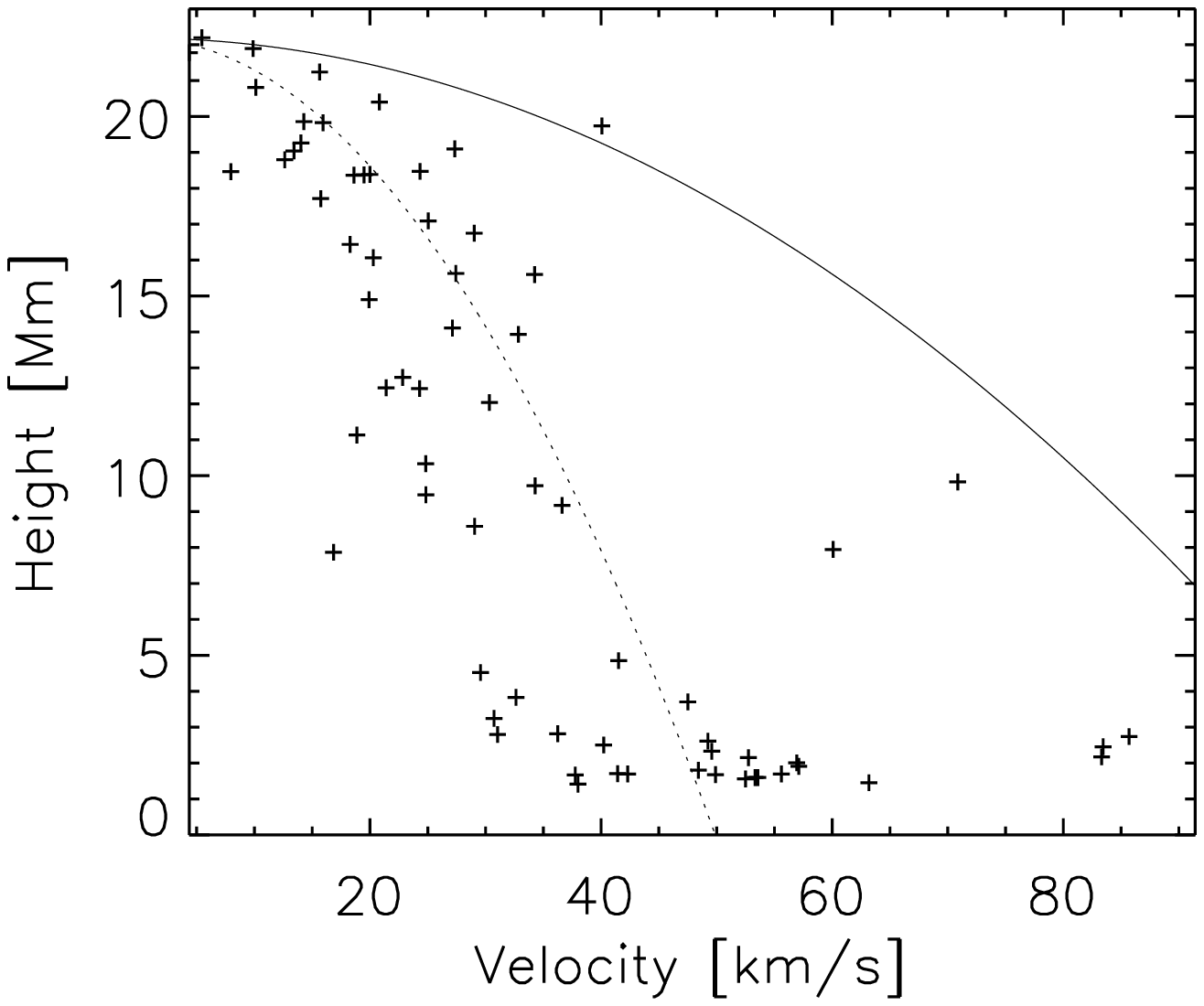}   \\
\end{array}$
\caption{Histograms of velocity (\textit{left}), acceleration (\textit{middle}) and (projected) height versus velocity (\textit{right}) for the coronal rain observed with \textit{Hinode}/SOT.}
\label{fig3}
\end{center}
\end{figure}

The condensations that can be tracked from high up in the corona down to chromospheric heights normally have a considerable thickness of about half a megameter. However, separation and elongation of the condensations generally occurs during the fall, leading to very thin and elongated condensations tracing strand-like structures. Many smaller condensations can be observed at various heights but are normally too faint to be followed clearly along their paths. In the observed loop a total of 28 condensations can be easily tracked to the chromosphere. Velocities and accelerations are derived with the help of length-time diagrams, which are shown in the left and middle panels of Figure \ref{fig3}. Since the velocity of the condensations varies along their paths, when possible, multiple velocity measurements at different heights are made, which allows to estimate the acceleration at different heights. The statistics in this loop are similar to those of the other loops on the north side of the sunspot. A broad distribution of velocities between 20~km~s$^{-1}$ and 100~km~s$^{-1}$ with a mean below 40 km s$^{-1}$ is obtained. The accelerations have in average lower values and are concentrated around a mean of 0.056~km~s$^{-2}$. Now, the change of the average effective gravity for a blob in free fall from the top of an ellipse with respect to its ellipticity can be calculated easily as $\langle g_{eff}\rangle=\frac{2}{\pi}\int_{0}^{\pi/2}g_{\sun}\cos\theta(s) ds$, where $\theta(s)$ is the angle between the vertical and the tangent to the path and $s$ is a variable parametrizing the path. It is found that for a ratio of loop height to half baseline between 0.5 and 2, $\langle g_{eff}\rangle$ varies roughly between 0.132~km~s$^{-2}$ and 0.21~km~s$^{-2}$, values that are significantly larger than the observed average value. A few cases of faster acceleration, as well as decelerations and constant falling velocities are also observed. In the right panel of Figure \ref{fig3} we have plotted the heights of the measurements for the condensations with respect to their velocity at that height. The solid and dashed lines in the figure correspond to the free fall speed under the action of the solar surface gravity and with the mean observed acceleration, respectively. 

Since Doppler velocities are not available in the cases analyzed in this paper we can only measure projected velocities without further assumptions on the geometry of the loop. The velocities and accelerations in the panels are thus lower estimates of the true values. We can however make an estimate of the errors. From Figure \ref{fig2} the projected distance on the plane of the sky between the two footpoints of the loop is estimated to be $l \simeq12$ Mm. This implies an angle between the line of sight and the plane of the loop of roughly $14^{\circ}$. Naming $h$ and $H$ the height of the measurement and the total height of the loop, respectively, the obtained error for each measurement results

\begin{equation}\label{error}
\mathrm{error} = \frac{hl/2H}{\sqrt{1-(hl/2H)^{2}}}.
\end{equation}

Since the upper and lower heights of the measurements are around 17 Mm and 2 Mm respectively, Equation (\ref{error}) gives an error of $16.5\%$ and $2\%$ for the upper and lower velocity measurements, respectively. This results in an error = $12.5\%$ for the acceleration. Hence the true mean acceleration is roughly 0.049 km s$^{-2}$. 

\subsection{Oscillations}\label{oscill}

\begin{figure*}
\epsscale{1.}
\plotone{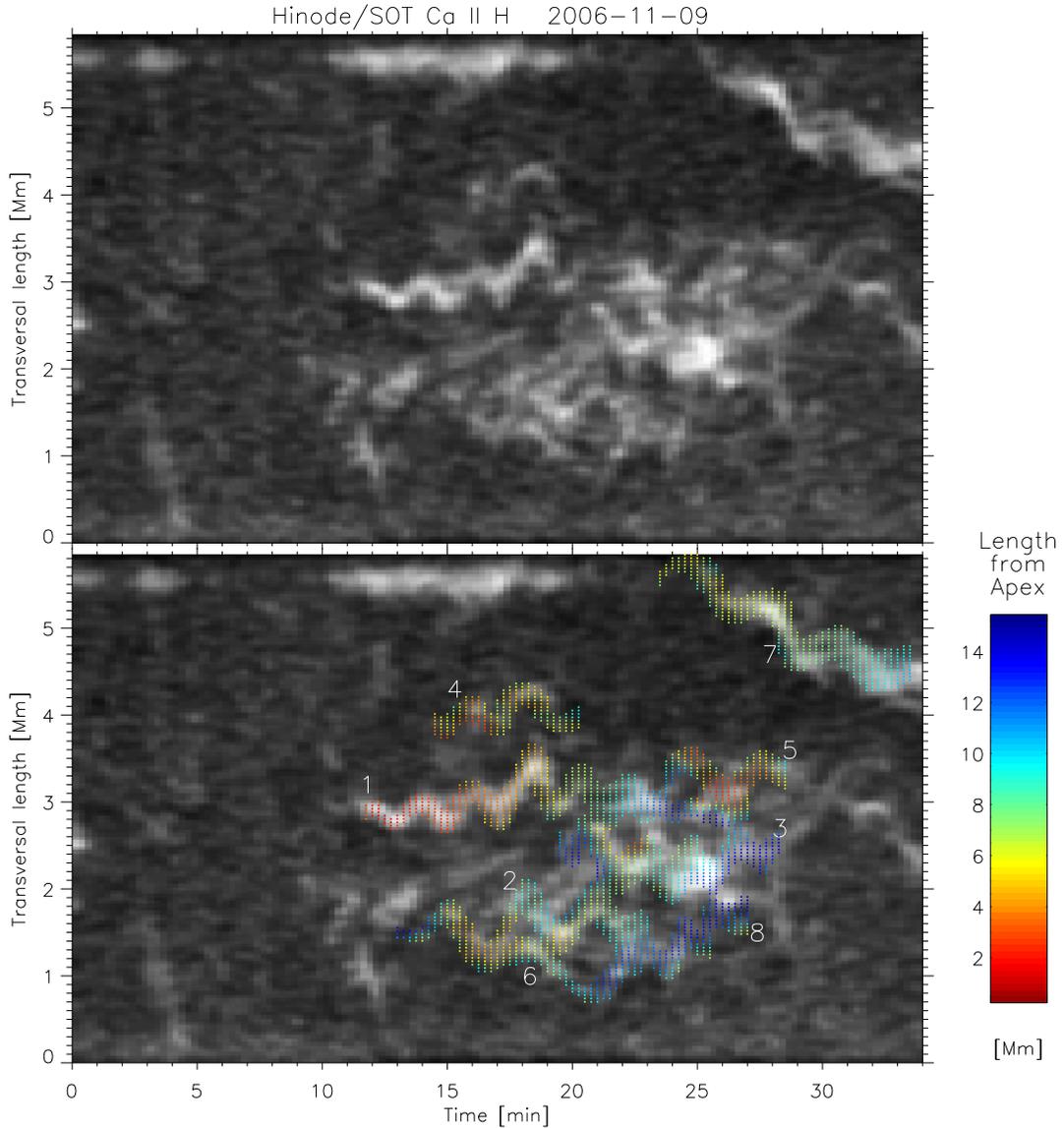}
\caption{Time slice across the loop. The transversal width corresponds to the distance between the two dashed lines in Figure \ref{fig2}, along a perpendicular to the dotted line. The time interval includes the time when coronal rain is observed. 8 oscillations can be detected. We repeat the figure, plotting in color over the oscillations the length from the apex where they occur.
\label{fig4}}
\end{figure*}

As the condensations fall, separation and elongation processes occur, which results in several strand-like structures resolved in the loop. The strands are observed to oscillate transversally. Figure \ref{fig4} shows the time slice along the outlined loop of Figure \ref{fig2}, where the transverse length refers to the perpendicular distance to the dotted line in Figure \ref{fig2}, from dashed line to dashed line. 8 oscillation patterns can be clearly observed, over which we plot in color the (projected) distance where they happen from the apex of the loop. There is a clear in-phase oscillation for the strands 1, 2, 3, 4 and 6 between the time period [15, 22] min in the figure. These strands become all visible simultaneously at roughly 8 Mm from the apex. 

\input{table1}

In Table~\ref{table1} the estimated periods, peak to peak amplitudes and transversal displacement velocities of the oscillations are shown. The periods lie between 100 s and 200 s, amplitudes are all roughly below 500 km and transversal velocities are between 4~km~s$^{-1}$ and 8~km~s$^{-1}$. The standard deviations are calculated by taking into account the widths of the strand-like condensations and the possible errors that they involve. Since these widths can be up to 500 km, this leads to large uncertainties in the oscillation amplitude measurements. This is also reflected in the calculation of the distance from the apex where the oscillation occurs (plotted in color in Figure~\ref{fig4}), where the distance does not always vary smoothly. 

Along their paths from the apex towards the chromosphere, the strand-like condensations reach a maximum separation between each other at an apparent distance of 4 Mm to 8 Mm from the apex of the loop, after which they gradually converge in the lower part of the loop's leg to a common footpoint about 1 Mm wide in the chromosphere. The maximum observed separation between the strands is close to 5 Mm wide. If the condensations indeed follow the magnetic field, this implies a magnetic geometry with a cross-sectional area expansion factor of at least 25 in 20 Mm height between chromosphere and corona. 

In the present observations by \textit{Hinode}/SOT several strand-like structures in a loop are observed to oscillate in-phase (synchronous). This points to either one transverse magnetohydrodynamic wave that affects all blobs as part of one monolithic loop, or either multiple transverse waves excited in separate strands but excited in phase by a common large-scale source. Possibly, the oscillations are not only confined to the specific structures but can involve the larger coronal region, and thus could be related to the prominence oscillations visible in the background \citep{Okamoto_2007Sci...318.1577O}. However, although the oscillation periods are similar, the difference in mode polarisation and the absence of oscillations in loops at the north side of the sunspot imply that the line-of-sight distance to the prominence may be too large to expect a common excitor.

\begin{figure}
\hspace*{-0.5in}\epsscale{1.4}
\plotone{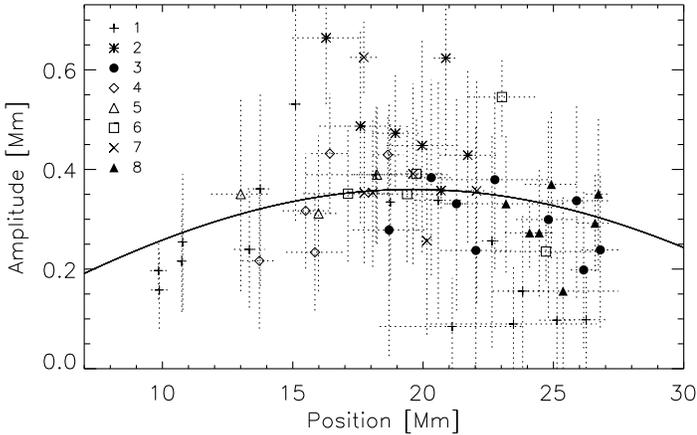}
\caption{Oscillation amplitude (transversal displacement from peak to peak oscillation) with respect to position along the loop. The position is calculated from the projected length from the apex (0 is at the apex), assuming a circular loop of height 25 Mm. The calculated points correspond to averages over boxed regions around each peak to account for the width of the strand. The error bars in position correspond to the standard deviations. The error bars in amplitude correspond to the width of the strand at each oscillation peak. The solid line corresponds to a fitted to the data of the first harmonic sine profile.
\label{fig5}}
\end{figure}

Information about the longitudinal and propagatory nature of the wave producing the observed oscillations can be obtained by analyzing the change of the oscillation amplitudes with respect to the position along the loop. This is plotted in Figure \ref{fig5}. Each point corresponds to a peak-to-peak amplitude of a specific oscillating strand and the corresponding position along the loop where it happens, corrected for projection effects assuming a circular loop of 25 Mm in height. The error bars in position correspond to the standard deviation of all the calculated positions in a boxed region around each peak. The error bars in amplitude are set equal to the width of the strand at each oscillation peak, which explains why they are large.

Since the oscillations in the loop can only be observed when the condensations are falling it is difficult to directly ascertain whether the agent is a propagating or a standing wave. Figure \ref{fig5} shows that the condensations do not appear to be oscillating when they are at the apex of the loop, nor in the lower part of the loop close to the footpoint. Furthermore, the amplitudes indicate a maximum at roughly half way along the loop leg (one fourth of the total loop length), which correspond to signatures of the first harmonic of a standing mode. For comparison, the solid line in the figure corresponds to a fitted sine profile to the data, which is the profile that a first harmonic would have. Alternatively, one could envisage a propagating wave packet, propagating up or down, the maximum amplitude meeting the condensations half way through the loop's leg. However, it is difficult to see how a propagating wave may reach a maximum amplitude at a given height independent of wave amplitude. Therefore, this scenario is less likely.

Apart from creating a transversal oscillation of the condensations in the loop, the observed waves may also have an effect on the general dynamics of coronal rain. As seen in the middle panel of Figure \ref{fig3} the observed accelerations do not have a broad distribution but concentrate around a low mean value of 0.056~km~s$^{-2}$. The range of the distribution is considerably smaller than that of the coronal rain observed in the coronal loops on the north side of the SOT field of view \citep{Antolin_2010ApJ...716..154A}. This may be a signature of a specific force acting in the upward direction. A net upward wave pressure should be present, which could be both in the (upward) propagating wave scenario or in the standing wave scenario. In the later, a first harmonic would produce in the upper first half of the leg a downward acceleration, followed by a deceleration in the lower part of the leg, which is the portion of the loop that is mostly observed. A transverse magnetohydrodynamic wave would exert an average acceleration on the plasma proportional to $(8\pi\rho)^{-1}\Delta B_{\perp}^{2}/\Delta h$, where $B_{\perp}$ is the transversal component of the magnetic field, $\rho$ is the density of the plasma and $h$ refers to a particular height. The average value for the effective gravity along the loop that a condensation would feel is $2g_{\sun}/\pi=0.174$~km~s$^{-2}$, assuming a circular loop, which implies an observed average deceleration of $\simeq0.118$~km~s$^{-2}$. In order to obtain the observed deceleration purely by means of the wave pressure, assuming a typically high coronal number density of $3\times10^{9}$~cm$^{-3}$ for active region loops subject to catastrophic cooling \citep{Antolin_2010ApJ...716..154A}, we would need a variation of $\Delta B_{\perp}\simeq0.4$~G in a 1~Mm height difference. 1.5 dimensional MHD simulations of both, standing torsional Alfv\'en waves heating coronal loops, and propagating Alfv\'en waves powering the solar wind, typically exhibit such gradients in the azimuthal magnetic field component \citep{Antolin_2010ApJ...712..494A, Suzuki_2006JGRA..11106101S}. Furthermore, analytical analysis and numerical modeling by \citet{Terradas_Ofman_2004ApJ...610..523T} have shown that the ponderomotive force from MHD waves (specifically from standing waves) can create flows and lead to significant density enhancements at locations of maximum amplitude.

\input{table2}

\section{Discussion}\label{discussion}

\begin{figure*}[!ht]
\epsscale{1.}
\plotone{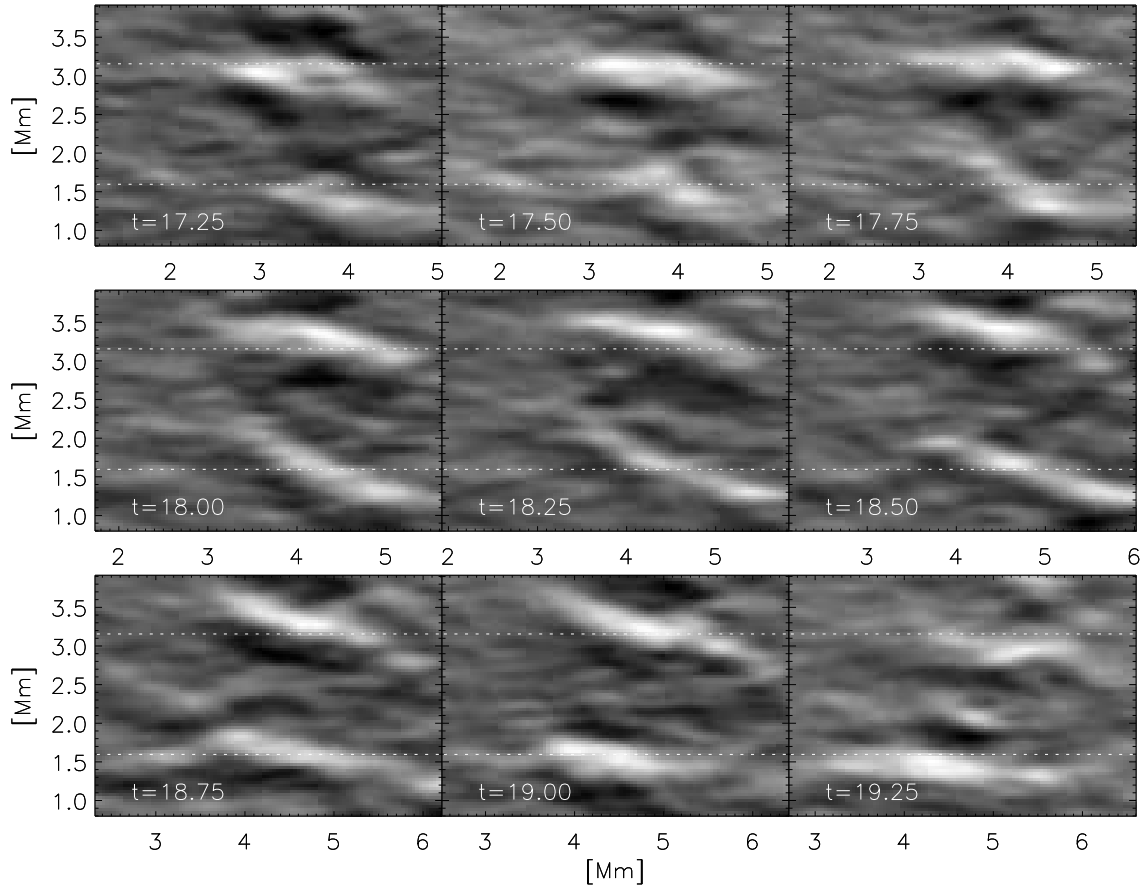}
\caption{Tracking of condensations 1 (upper) and 2 (lower) of Figure \ref{fig4}. The length in the x axis corresponds to the (projected) length from the apex of the loop. The length in the y axis corresponds to the same transversal length as in Figure \ref{fig4}. The times (in minutes) are set in the low left corner of each panel. Notice that each condensation does not oscillate up and down as a single structure but exhibits a different oscillation amplitude along its length.
\label{fig6}}
\end{figure*}

During the time interval in which the blobs are observed no phase shift is detected in the oscillations, reinforcing the standing wave scenario. In case that propagating waves are playing a role in the oscillation of the blobs, these have to propagate with a phase speed larger than that set by the uncertainty in the measurements. Taking for instance the case of strand 1, which is observed over a distance of about 12 Mm according to Figure~\ref{fig4}, with an uncertainty of 1 time step we have a lower bound for the phase speed on the order of 800~km~s$^{-1}$. For the onset of a first harmonic in the loop, the total wavelength is equal to the loop length, leading to phase speeds between 400~km~s$^{-1}$ and 1000~km~s$^{-1}$ (see Table \ref{table2}). To give a measure of the energy contents of these waves, we calculate the wave energy flux for the case of strand 1 as would be if the wave was propagating with half the amplitude. For this case, the phase speed is $v_{ph}=760\pm264$~km~s$^{-1}$, giving an Alfv\'en speed of $v_{A}=v_{ph}/2^{1/2}=540\pm186$~km~s$^{-1}$. The energy flux is then given by $E_{\textrm{flux}}=\frac{1}{2}\rho v_{t}^{2}v_{A}$, where $\rho=\mu n_{e}m_{p}$ is the average coronal density of the loop (taking an electron number density of $n_{e}=3\times10^{9}$~cm$^{-3}$ and $\mu=1.27$ the effective particle mass with respect to the proton mass $m_{p}$), and $v_{t}$ is the transversal velocity calculated in Table \ref{table1}. Replacing with the numerical values we obtain an energy flux of ($4.9\pm5.9$)$\times10^{4}$~erg~cm$^{-2}$~s$^{-1}$, where the large uncertainties are due to the large standard deviations in the observed periods and transversal velocities. The calculated value is below the estimated necessary energy of $10^{6}$~erg~cm$^{-2}$~s$^{-1}$ for heating the active region corona \citep{Withbroe_1977ARAA..15..363W}. In order to assess how much energy is transferred from the standing wave to the plasma, information about the wave damping is necessary. However, in our case, it is difficult to find conclusive evidence for damping of the oscillation because in tracking a falling condensation the information of the height-dependence of the oscillation profile and the time-evolution are intertwined.

\begin{figure*}[!ht]
\epsscale{1.1}
\plotone{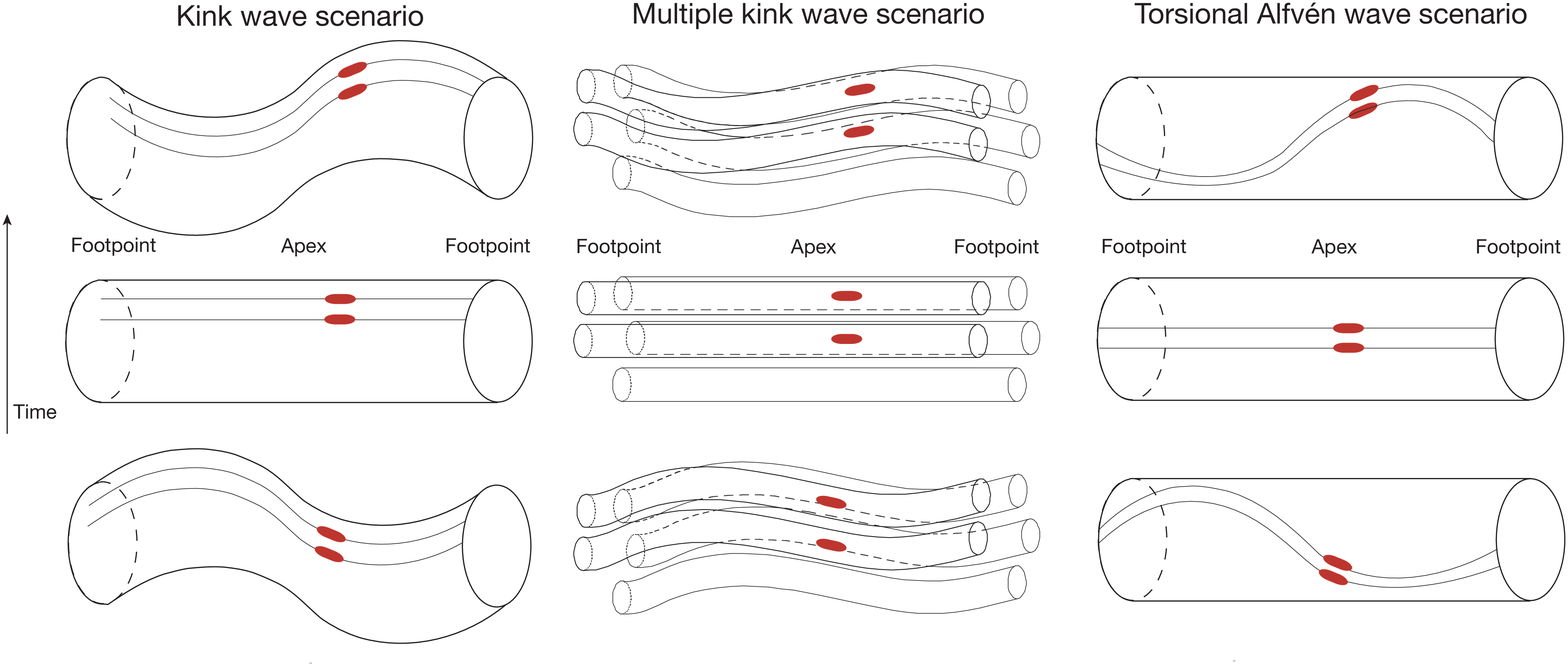}
\caption{Sketch illustrating the possible interpretations to the oscillatory events. First, a standing kink mode corresponding to a first harmonic, that displaces the axis of a single monolithic wave guide occupying the whole magnetic structure (\textit{left}). Second, multiple standing kink modes, each within individual strands in the magnetic structure guided by local density enhancements associated with each blob (\textit{middle}). Third, a torsional Alfv\'en wave that occupies one or several flux surfaces within the single monolithic loop producing a swaying of the individual strands from the twisting motions, without displacing the main axis of the loop (\textit{right}). The blobs in red represent the condensations falling along the loop, from the apex towards the right footpoint. 
\label{fig7}}
\end{figure*}

Since we are observing the loop in a state of cooling, long after the heating has taken place, it is possible that the initial oscillations were faster, thus implying a much larger energy flux for the initial propagating waves setting up the standing oscillations. Let us assume this scenario, and a sufficient energy flux to heat the corona of $10^{6}$~erg~cm$^{-2}$~s$^{-1}$ at the beginning of the 'condensation' cycle (the cycle where the catastrophic cooling takes place, also known as 'limit cycle'). An initial coronal electron number density of $10^{9}$~cm$^{-3}$ (a rough average for active region loops) and a coronal magnetic field of $\lesssim$10~G \citep[taking Equation \ref{Bkink}; see also Figure 12 in][]{Nakariakov_2005LRSP....2....3N} leads to a transversal velocity for the oscillations of at least 22~km~s$^{-1}$. This means that in a time interval of about 20 to 30 minutes \citep[which is an estimation of the 'condensing' phase time prior to catastrophic cooling, based on results of][]{Antolin_2010ApJ...716..154A} our oscillations have slowed down to about 1/3 to 1/4 their initial values at the beginning of the cycle. During the catastrophic cooling we do not observe any damping (mostly due to the fact that we cannot observe the same portion of the loop over a significant time interval), but it is possible that the main damping has occurred already during the condensing phase. As decaying mechanisms, resonant absorption has been shown to be an effective mechanism \citep{Ruderman_2002ApJ...577..475R, Goossens_2002AA...394L..39G}. Assuming an exponentially decaying amplitude, and taking an observed mean transversal velocity of 5~km~s$^{-1}$ (according to Table \ref{table1}), we obtain a damping rate of (0.05 - 0.075)~min$^{-1}$ for a time interval of (20 - 30)~min. Thus leading to a decaying time of (13.5 - 20)~min, which is in the order of whatÊhas been reported (matching the periods observed here) in similar loops \citep{Nakariakov_1999Sci...285..862N, Aschwanden_2002SoPh..206...99A}. This calculation is however strongly dependent on the time of the condensation cycles (limit cycles), which are still subject to debate. To confirm such scenarios direct evidence from EUV imagers in future events will be required. On the other hand, if the initial energy of the wave is not dissipated but conserved, it is worth noting that the transversal velocities of the oscillation are not altered significantly by an increase of density through the condensation processes, since it can be seen from conservation of wave energy flux that the transversal velocity amplitude $v_{t}$ depends only weakly on density increases as $\rho$ as $v_{t} \propto \rho^{-1/4}$.

The determination of the local magnetic field seismologically depends on the correct identification of the wave. This, in turn, depends on the existence of a waveguide in the loop. Prior to a catastrophic cooling event, simulations show that the loop undergoes a phase of increasing coronal density and slowly decreasing coronal temperature \citep{Antiochos_1999ApJ...512..985A, Muller_2004AA...424..289M, Antolin_2010ApJ...716..154A}. Figure \ref{fig1}, which corresponds to an XRT observation roughly half an hour before the observed coronal rain, shows that most of the region in the SOT field of view is composed by plasma with temperatures above a million degrees. Our loop is not directly visible, which may be due to the cooling phase prior to the catastrophic cooling event. During this phase, which constitutes most of the cycle, the average coronal density reaches values that are considerably higher than that of the average exterior corona. According to simulations, this dense loop state, which can constitute a waveguide, is maintained for times on the order of tens of minutes, enough for the onset of standing modes from fast magnetoacoustic waves. In our case, the observed transverse displacement can be associated with one of three types of waves, sketched in Figure \ref{fig7}. First, the wave is a kink mode that displaces the axis of a single monolithic wave guide that occupies the whole magnetic structure. Second, there may be multiple kink modes, each within individual strands in the magnetic structure guided by local density enhancements associated with each blob. Third, the wave is a torsional Alfv\'en wave that occupies one or several flux surfaces within the single monolithic loop. The main axis of the loop is then not displaced but we observe the swaying of the strands from the twists or torsional motions within flux surfaces. A slow magnetoacoustic wave is excluded from consideration because it is essentially a longitudinal mode.

For all cases the average coronal magnetic field in the loop can be determined from:
\begin{equation}\label{Bkink}
B_{0}=\sqrt{2\pi}\frac{L}{P}\sqrt{\rho_{0}(1+\frac{\rho_{e}}{\rho_{0}})}
\end{equation}
\citep[formula 31 in][corrected for the first harmonic]{Nakariakov_2005LRSP....2....3N}, where $\rho_{e}$ and $\rho_{0}$ are the exterior and interior densities respectively. This equation is valid under the assumption that the width of the loop is much smaller than the total length, which holds according to the estimates of section \ref{oscill}. The case of $\rho_{e}$=0 corresponds to the lower limiting case for a kink mode, whilst $\rho_{e}$=$\rho_{0}$ would correspond to the case of a torsional Alfv\'en wave. In Table \ref{table2} we have estimated the magnetic field in both cases, assuming the height of the loop to be 25~Mm and a loop number density of $3\times10^{9}$ cm$^{-3}$, a normal value of dense loops subject to catastrophic cooling \citep[our loop here is not so different from the modeled loop in][]{Antolin_2010ApJ...716..154A}. Taking the two limiting cases of exterior-to-inside density ratios of 0 (kink mode scenario) and 1 (torsional Alfv\'en wave scenario) we obtain an average coronal magnetic field of 11.5~G and 16.3~G, respectively. Importantly, when applying Equation \ref{Bkink} we make the assumption that the blobs themselves do not carry sufficient inertia to affect the oscillation itself. A scenario in which this assumption does not hold is worth investigating, and will be addressed in a subsequent paper.

It is interesting to note that the collective behavior of the strands in the loop does not seem to last for the entire falling time of the coronal rain. Figure \ref{fig4} shows that after roughly $t=21$~minutes it is difficult to observe any synchronous oscillation of the strands. This loss of collective behavior may be apparent, resulting from several factors such as, a change in the magnetic field geometry of the lines, a change of the morphology of the condensations (increasing the difficulty of differentiating between the strands), or a change of the plasma conditions in the condensation (reheating or further cooling, or a thinning of the plasma leading to a change in the opacity in the \ion{Ca}{2} H line). If the loss of collective behavior is however real, it would exclude the option of a kink mode in a single monolithic loop as it should remain collective with the loop  \citep{VanDoorsselaere_2008ApJ...676L..73V}. For the other two options, i.e. kink modes in loop strands and a torsional Alfv\'en wave in a single loop, the loss of collective behavior can be explained in terms of phase-mixing. For the former option, variations in the density of the different strands leads to a drifting out of phase of the individual blob oscillations. For the latter option, if the torsional modes belong to different magnetic surfaces inside the loop, different Alfv\'en speeds are expected, leading in turn to a loss of phase. This could explain, in turn, the obtained different periods in Table \ref{table1} and different phase speeds in Table \ref{table2}. 

\citet{Terradas_etal_2008ApJ...678L.153T} have recently shown analytically that a standing kink mode in a loop where siphon flows are present will experience a linear shift in phase with position along the loop and an asymmetric profile in time of the eigenfunctions with respect to the loop's apex. The short interval of time during which the loop can be observed does not allow us to detect any overall phase shift for the observed strands. However, due to the relatively low speed of the detected flow ($\lesssim60$~km~s$^{-1}$), we do not expect its effect on the phase speed of the waves to be significant. For instance, if our oscillations correspond to a shifted fundamental mode we would still be able to observe the coronal rain oscillating significantly at the apex for several periods, which is not observed.

Figure \ref{fig6} shows a zoom-in of condensations 1 and 2 (and 6) while they fall. The $x$-axis corresponds to the projected distance from the apex of the loop (coded in color in Figure \ref{fig4}), and the $y$-axis corresponds to the same transversal width across the loop as in Figure \ref{fig4}. The corresponding times are set in the lower left corner of each panel. The 2 minute sequence corresponds approximately to one period of strands 1 and 2, when these exhibit the maximum oscillation amplitude. As can be seen in the Figure, both condensations start horizontal (time $t=17.25$ min), are slanted halfway through the oscillation ($t=18.25$ min), and end up again horizontal at the end ($t=19.25$ min). Under the assumption that the condensations retain their shape during the 2 minutes of the sequence, so that we are indeed following the same plasma parcel, the later means that the condensations do not oscillate up and down uniformly but rather exhibit amplitude differences along their lengths. As shown in the sketch of Figure \ref{fig7}, in the scenarios offered by both a kink mode and a torsional Alfv\'en wave it is hard to explain such effect only on the basis of, respectively, the displacement of the loop axis and the twisting of the magnetic field lines. This is due to the fact that the observed wavelength of $\simeq80$~Mm (in case of a first harmonic) is much longer than the length of the condensations ($\lesssim3$~Mm), and also to the fact that the phase speed of the wave ($\gtrsim400$~km~s$^{-1}$) is much faster than the falling speed of the condensation ($\lesssim60$~km~s$^{-1}$), and hence we should expect the whole condensation to displace transversally in a uniform way along its length. In the case that the length of the condensation is not so short when compared to the wavelength (as in the sketch of Figure \ref{fig7}, we would expect a periodical change of the slope of the condensations (with respect to the axis of the loop), which is not observed in our case. If we have instead a propagating wave, different portions of a condensation could oscillate with different amplitudes at a given time, but eventually all portions of the condensation should exhibit the same maximum amplitude, which is not observed neither. Hence this effect may not be caused by the nature of the oscillations.

We believe the cause for the observed effect in Figure \ref{fig6} to be linked to the cause of the often observed separation (and subsequent elongation) of the plasma in the condensation. Due to the high density of the condensation it is possible that the plasma beta parameter is high enough that the plasma moves transversally to the axis of the loop, thus allowing also the observed separation process. The density and temperature range of coronal rain are not well known observationally, but since its opacity is large enough to appear bright and dark in H$\alpha$ and \ion{Ca}{2}~H when observed above limb and on disk (towards the limb) respectively \citep{Antolin_etal_11}, the range of values must be chromospheric. This is also supported by numerical simulations of catastrophic cooling \citep{Muller_2003AA...411..605M, Muller_2004AA...424..289M, Antolin_2010ApJ...716..154A}. In the later the temperature of coronal rain is estimated to be as low as $6\times10^{4}$~K and its number density to be about $10^{11}$~cm$^{-3}$. Taking a coronal magnetic field inside the loop of $14$~G, an average of the values calculated in Table \ref{table2}, we obtain a rough estimate for the plasma beta parameter of $\beta=8\pi p/B^{2}\simeq 8\pi n k_{B}T/B^{2}\simeq0.1$. Since the loss of thermal equilibrium implies high velocity upflows and subsequent shocks, it is not unreasonable to consider the possibility of the plasma beta parameter becoming high enough so that the plasma expands transversally. 3D numerical simulations of coronal rain formation are needed to correctly address this idea, which is the subject of a future work.

After separation of the initially dense condensation, the plasma in coronal rain is observed to elongate into strand-like structures, thus probably tracing the internal structure of coronal loops. Whether the elongation process is just a result of gravity acting differentially along the magnetic topology or if other more sophisticated processes are involved is an interesting question that needs to be addressed properly with the help of numerical simulations. For instance, since the region below the falling blob is expected to be magnetically dominated, according to the previous discussion the layer in between should meet the criterium for the onsetting of the magnetic Rayleigh-Taylor instability, which would contribute to the separation of the blob. This effect is thought to be responsible for the finger-like structures observed in prominences \citep{Berger_etal_2010ApJ...716.1288B}.
 
The causes for the observed oscillations are less clear. Reported horizontal oscillations of loops are often linked directly or indirectly to flares or CMEs \citep{Nakariakov_etal_2009AA...502..661N}. However, in our observations no flares or energetic events were reported on that day. A possible cause may be interaction with neighboring loops. In Figure~\ref{fig0} we have outlined in dotted curves some paths of coronal rain marking the presence of other loops. Due to the projection effect it is hard to say whether these loops are indeed close-by and whether any interaction really occurs. It is interesting to note however that the coronal rain in these loops occurs prior to the coronal rain in our loop, and that the paths seem to intersect roughly halfway through the visible leg of our loop. If this is not a projection effect, it is possible that the coronal rain perturbs our loop thus producing the oscillations. The perturbation would have a maximum amplitude at the crossing region, thus explaining why we observe a maximum amplitude for the oscillations halfway along the leg.

Another possibility can be a scenario in which the inertia of the condensations conforming coronal rain is not negligible, thus affecting the stability of the entire loop (and hence triggering themselves oscillations in the loop). Making a rough analogy to a hose with water gushing in, it is natural to expect that there might be a limit for the quantity and for the velocity of the plasma above which the stability of the magnetic field structure is compromised and the loop oscillates. This scenario is worth investigating through numerical simulations, and will be the subject of a subsequent paper.

An additional possible scenario for triggering transversal oscillations in loops is one in which the waves are linked to an unobserved energetic event such as magnetic reconnection in the lower atmosphere, triggered by convection. 3D Numerical simulations have shown that magnetic reconnection processes in the chromosphere (in a solar atmosphere powered by convection) can easily generate energetic events such as spicules and other jet-like phenomena \citep{Martinez-Sykora_2009ApJ...701.1569M, Martinez-Sykora_2010arXiv1011.4703M, Heggland_etal_2011}. It is thus reasonable to consider reconnection as a possible candidate to generate transverse perturbations of flux tubes. On the other hand, swirl events have been observed in the photosphere  \citep{Wedemeyer-Bohm_2009AA...507L...9W}. Numerical simulations have shown that torsional motions at photospheric level can generate different kinds of modes in flux tubes \cite{Fedun_etal_2011, Shelyag_etal_2011b} and input enough amounts of energy for coronal heating \citep{Kudoh_1999ApJ...514..493K, Antolin_2010ApJ...712..494A}.

Since we expect a certain degree of twisting and braiding for the strands in the observed loop, a last possible scenario we can think of is one in which the observed oscillatory events are not due to waves, but result from the internal complex topology of the loop. The condensations would then fall down the braided 'helical' strands which would make it seem they are oscillating. We consider however this scenario to be unlikely. Since we observe the oscillations over several strands, and mostly in-phase, the twist in the loop would need to exist over most of the loop's width. Also, we would need the tube to be twisted along its entire length in order for it to be in equilibrium in the corona. However, the twist we observe seems to have 2 nodes, one at the apex and one towards the footpoint, and its amplitude increases in between. Furthermore, the stability of such a configuration has to be maintained over the time interval in which the oscillatory events are observed, which is longer than 20 minutes.

\section{Conclusions}\label{conclusions}

We have analyzed transversal oscillations of a loop that are put in evidence by coronal rain falling in the loop. The coronal rain is observed to fall down from the apex, roughly $25\pm5$~Mm above the surface. The condensations composing coronal rain are observed to separate and elongate as they fall down, exhibiting large distributions of velocities but rather concentrated accelerations around a much lower value than that of the average effective solar gravity along the loop, thus implying the presence of a different force, probably of magnetic origin. The obtained deceleration can be the result of wave pressure from transverse magnetohydrodynamic waves at coronal heights.

As the condensations fall, they elongate into strand-like structures that are observed to oscillate in-phase transversally to the axis of the loop with periods that are similar to those normally observed in prominences. The amplitudes of the oscillations are observed to vary significantly with respect to the position along the loop, having a maximum at roughly halfway through one leg and minimums at both the apex and towards the footpoint of the leg. We have interpreted this result as a signature of a first harmonic of a standing transverse magnetohydrodynamic wave in the loop, although an upward propagating wave is also a possible, but less likely, scenario. This interpretation implies a wavelength equal to the loop's length, $80\pm15$~Mm. 

Since this active region loop exhibits a catastrophic cooling event we expect the internal density to be significantly higher than that of the external corona for an interval of time long enough to create a waveguide along the loop. The obtained phase speeds of the waves are between 400~km~s$^{-1}$ and 1000~km~s$^{-1}$, implying either a fast (horizontal) kink mode or a torsional Alfv\'en mode. The wide distribution of speeds may be due to the possible uncertainties in the measurements, given the short time in which the loop can be observed. On the other hand, if the distribution in the phase speeds is real, it implies a loss of collective behavior which can be explained in terms of phase-mixing, present both in a scenario in which each strand has its own kink mode, and in the scenario of a torsional Alfv\'en wave. Recently, \citet{Harris_etal_2011} reported collective behavior in transverse oscillations of an ensemble of prominence threads. There also the collective nature is lost with time as the threads oscillate with slightly different periods and amplitudes. The average coronal magnetic field inside the loop is estimated to be between $8\pm2$~G and $22\pm7$~G, in agreement with other estimates of coronal magnetic fields in active region loops through coronal seismology techniques \citep{Nakariakov_Ofman_2001AA...372L..53N}. 

The strand-like condensations do not exhibit a uniform oscillation amplitude along their lengths, which we believe is not caused by the oscillation, but by the physical conditions inside the loop. A rough estimate indicates an average plasma-$\beta$ parameter of 0.1 in the condensation due to the high densities and strong shocks that are normally created by the catastrophic cooling mechanism. This allows us to interpret the result as evidence of strong gas pressure forces relative to magnetic forces inside the loop, which would explain the often observed initial separation of the condensations and which can also be linked to the deceleration of the plasma.

Coronal rain has been previously shown to be deeply linked with the coronal heating mechanism in the loop. Here we have shown the potential it can play in understanding the magnetic field topology of the solar corona by tracing the internal substructures of loops, marking the internal forces at play, exposing wave-like phenomena and thus allow the measurement of the coronal magnetic field strength by means of coronal seismology. We have pointed to several important problems that need to be addressed in future work. Namely, by means of 3D numerical simulations, investigation of the possible processes allowing the separation and elongation of the condensations in the coronal rain, investigate the internal physical conditions of such condensations created through the catastrophic cooling mechanism by means of a proper radiative transfer model of the atmosphere, and investigate coronal rain being itself a possible cause of transverse magnetohydrodynamic oscillations of loops. 

\acknowledgments
This work was supported by the Research Council of Norway and the UK Science and Technology Facilities Council through the CFSA-Warwick Rolling Grant. Hinode is a Japanese mission developed and launched by ISAS/JAXA, with NAOJ as domestic partner and NASA and STFC (UK) as international partners. It is operated by these agencies in co-operation with ESA and NSC (Norway). The authors would like to thank Mats Carlsson, Val\'ery Nakariakov and Claire Foullon for the helpful discussions that led to a significant improvement of this manuscript, and to Tom Van Doorsselaere for having first detected the oscillatory event with his sharp eyes and directing us to it. We would also like to thank Gregal Vissers for the development of the splendid CRISPEX tool, and the referee for the constructive comments. P. A. would like to acknowledge Siew Fong Chen for artistic support and patient encouragement. 

\bibliographystyle{aa}
\bibliography{antolin_verwichte_2011_preprint.bbl}  

\end{document}